\documentclass[aps,twocolumn,preprintnumbers,amsmath,amssymb,floatfix,groupedaddress,nofootinbib]{revtex4}

\usepackage{enumerate}
\usepackage{graphicx}
\usepackage{color}
\usepackage[utf8]{inputenc}
\usepackage{enumitem}
\usepackage{appendix}

\usepackage{fontenc}  
\usepackage{amsmath}

\begin{document}
%=================================================================
% Full title of the paper (Capitalized)
\title{Postulating the Unicity of the Macroscopic Physical World.}

\author{Mathias Van Den Bossche$^1$ and Philippe Grangier$^2$}

\affiliation{$^1$ Thales Alenia Space, 26, avenue J.-F. Champollion, 31037 Toulouse, France}
\affiliation{$^2$ Laboratoire Charles Fabry, IOGS, CNRS, Universit\'e Paris Saclay, F91127 Palaiseau, France.}

\begin{abstract}
We argue that a clear view on quantum mechanics is obtained by considering that the unicity of the macroscopic world is a fundamental postulate of physics,  rather than an issue that must be mathematically justified or demonstrated.  This postulate 
allows a framework in which  quantum mechanics can be constructed, in a complete mathematically consistent way.  This is made possible by using general operator algebras to extend the mathematical description of the physical world towards macroscopic systems. Such an approach goes beyond the usual type I operator algebras  
used in standard textbook quantum mechanics.  This avoids a major pitfall, which is the temptation to make the usual type I formalism `universal’. This may also provide a meta-framework for both classical and quantum physics, shedding a new light on ancient conceptual antagonisms, and clarifying the status of quantum objects.  Beyond  exploring remote corners of quantum physics, we expect these ideas to be helpful to better understand and develop quantum technologies. 

\end{abstract}

\maketitle

\section{Introduction}

\subsection{Our world is unique...}
%\vspace{-3 mm}
Obvious empirical evidence tells us that we live and die in a single world, which has an history before and after our individual existence. Alternative histories of the universe are possible but counterfactual, and remain a subject for fiction.  In a similar line of ideas, the past can be forgotten but not changed, and the future can be predicted and depends to some extent on our actions, or absence of action. It is actually a mix of quasi-certainty (the sun will raise tomorrow) and inherent uncertainties (it is likely to rain tomorrow). And once something has happened, good or bad, there is no way back.

These obvious empirical evidences have set the framework in which mankind has evolved, and they have basically not changed, despite the fact that our abilities to record the past, and to predict the future, have tremendously changed over the centuries and millenia. In the modern world there is no more need to kill Iphigenia, but propaganda is more active than ever.  Actually, most of our current techniques for recording, sending and processing {\color{black} --} also manipulating {\color{black} --} information are using microelectronics, that is based itself on quantum physics. 

Here we want to argue that considering
the unicity of the macroscopic world as a basic postulate for physics is not only possible, but appears as a firm basis on which quantum mechanics (QM) can be built.  So our approach  does not contradict  QM as it is currently used, but it embeds the usual quantum formalism in a framework where it is still valid, but where some misleading extrapolations don’t show up \cite{FL}. From the mathematical side this extended framework is not really new and may be traced back to John von Neumann at the end of the 1930’s~\cite{JvN39,onRingsOfOperators}. {\color{black} This framework} is also used in mathematical physics {\color{black} when addressing some aspects of} statistical physics and quantum field theory \cite{emch}. 
{\color{black} However 
it is conceptually and technically demanding, and has been mostly ignored  by physicists} in standard QM, and especially in elementary textbooks. But as we will see below it may  be extremely useful to establish the overall consistency of QM. 

\subsection{... but is it classical and/or quantum ?}
%\vspace{-2 mm}
 Since the definition of the theoretical framework of QM to describe the microscopic world initiated by Heisenberg and Schr\"odinger and completed by Dirac and von Neumann, the tremendous reliability of its predictions %it yields 
 combined with the exotic phenomena it unveils have raised much interrogation in the scientific community  and beyond. The exotic aspect has led to interpretations that many consider as far-fetched, and mainstream quantum physicists tend to stick to a 
 `don't bother and calculate!'  approach. We will call `textbook quantum mechanics' (TBQM) the formalism on which the mainstream approach relies \cite{CCT}. Despite its efficiency, this attitude may not be the one that will allow much future development -- especially in the frame of e.g. the emerging quantum technologies{\color{black},} as well as quantum gravity. 

The main problem that motivates non-mainstream interpretations lies in the non-deterministic aspect of quantum projective measurements. This aspect, when related to quantum superpositions,  leads to such wordings as `the system is in two states at the same time' in popular texts close to mainstream thoughts. More divergent schools consider that QM could be in fact deterministic at the cost of hidden mechanisms (Bohmian mechanics, pilot wave) or at the expense of a proliferation of replicas of the Universe every time a projective measurement happens (Everett's multiverse interpretation). It is true that the contrast between the smooth, predictable unitary evolution of quantum systems without measurement on the one hand, and on the other hand the abrupt, discontinuous, dissipative, random effect of a measurement, creates  shock waves that leave no-one indifferent.

A second problem lies in the significant difference between the laws that apply to microscopic vs. macroscopic systems,
all the more because macroscopic systems are made of a very large number of microscopic systems. This  leads to the ideas that macroscopical laws should emerge from laws of quantum physics as a sort of averaging out of quantum effects. This is the essence of a form of `reductionism', and it is true that Ehrenfest's theorem or the properties of coherent states seem to show at a path in this direction. Nevertheless another view, coming {\color{black} mostly from Bohr and Heisenberg} 
and adopted in TBQM,  tells that there is an irreducible cut between the microscopic and the macroscopic worlds.
TBQM can be considered as `dualist' when considered from this point of view. 

On the reductionist side, Zeh \cite{zeh} and Zurek \cite{zurek} have elaborated the concept of decoherence that explains how in the process of measurement, the microscopic degrees of freedom of the quantum system of interest become entangled with a huge number of external degrees of freedom (in the measurement device, in its environment). The global state cannot be realistically tracked thus leading to a leak of information outside the system. This leak can however be computationally managed in the density operator of the system and environment as a partial trace on the environment degrees of freedom. This  leads to a situation where this operator becomes diagonal on pairs of associated system and device states. Though this does not explain how a precise measurement result is selected among possible ones, at least it shows how system and device may couple to produce a result.

A third problem of a different nature has to be mentioned at this point. Our civilisation is on the verge of developing quantum technologies, which rely on using all the specific aspects of quantum physics, and not only some of them as in electronics and photonics, where a part of the {\color{black} weirdness}  is hidden by the large quantities of involved electrons or photons. These technologies will be developed by engineers who will be all the more efficient if they can develop an intuitive understanding of QM. Developing intuition requires clearing the various ad-hoc and often wobbly explanations that are given on such or such basic properties, often by awkwardly trying to connect them with a usual, classical view of the world. This means that clarifying `what exists' (ontology) and what happens whatever the observer (objectivity) is not only a philosophical question, it is the condition of progress of these technologies. 

The present article  supports the idea  that the `Contexts, System, Modalities'  approach (CSM) developed in previous papers by one of the authors and coworkers (see \cite{csm4c} and for more details \cite{CO2002,csm1,trsa,random,csm4b,myst,inference,debate,KS,completing,FoP2023,DICE2024,extWigner}) may provide  such a clarification to the understanding of QM.

The purpose of the present article is  first  to summarise the main features of CSM (section 2), that starts from a few empirical postulates, including the unicity of the physical macroscopic world, and the contextuality of measurements, in order to build up  the  formalism of TBQM. {\color{black} In order to present an overview of CSM, we will not give here  the details of the proofs, but refer to published papers. Some major steps in the CSM framework are to show that quantum}
theory must be probabilistic, and also that the description of macroscopic systems requires an {\it operator algebraic framework} broader that the one traditionally used in TBQM. This theoretical extension actually provides an understanding of the classical-quantum transition, and allows building a comprehensive overall framework to embed both realms. In section 3, we discuss several implications of this approach on a more general level, i.e, on the validity of considering mathematically infinite systems, and on  hazardous predictions related to the so-called ‘universal unitarity’. We will also argue that the reductionist vs. dualist views of  physics may not be so antagonistic after all. 

%%%%%%%%%%%%%%%%%%%%%%%%%%%%%%%%%%%%%%%%%%
\section{Overview of the reconstruction of non-fully-unitary QM. }

\subsection{Introduction and motivation.}

The heart of the conceptual difficulties of quantum mechanism lies in the measurement process, that is framed in TBQM by the von Neumann projection postulate. What is the  problem with this postulate ? Certainly not to be wrong, since it is at the core of the functioning of TBQM, and it has been observed correct whenever the relevant observations could be made. The main usual criticism is that it introduces a projection, that is a non-unitarity evolution during a measurement, in apparent contradiction with the fact that the system and measurement apparatus should globally evolve according to Schr\"odinger equation, predicting a unitary evolution. 

There are many ways to deal with this unsatisfactory dichotomy, most of them are based on the decoherence theory \cite{zeh,zurek}, telling basically that measurement devices are large systems coupled to an even larger environment, where interactions ultimately create an extremely large entangled system. Therefore it is not possible to keep track of all degrees of freedom, and unitarity is broken by taking a partial trace over the degrees of freedom that get out of control. In addition to this loss of information, the structure of the interaction with the environment selects the measurement basis by environment-induced selection, that is ‘einselection’ \cite{zurek}. 

After some reasonable approximations this leads to a classical probability distribution over the possible results in the einselected basis, but one more problem shows up: only one result is observed, not a probability distribution of them, so what 
does ‘select the winner’ in an individual measurement event ? This question is more tricky, because it assumes that there should be such a winner selection process, that does not exist in basic QM. So a simple way to avoid the problem is to tell that the result is fundamentally a probability distribution, and there is nothing like a  winner selection process. Then the unique result can simply be observed at the macroscopic level, and the probability distribution is actualized as usual. 
%\vspace{-2mm}
\subsection{The proposed reconstruction.}
%\vspace{-2mm}
Clearly this observation plus actualisation process is fully consistent with a postulate on the unicity of the physical macroscopic world, and thus of the macroscopic measurement result. The CSM framework \cite{csm4b,CO2002,csm1,trsa,random,myst,csm4c,inference,debate,KS,completing,FoP2023,DICE2024,extWigner} is built upon this idea, with the following postulates:
%\\

\begin{itemize}

\item[\bf P$_0$]{\bf Unicity of the macroscopic world} --
There is a unique macroscopic physical world, in which a given measurement gives a single result. 

\item[\bf D$_1$] {\bf Contexts, systems and modalities} --
Given a {\color{black}microscopic} physical system, a {\it modality} is defined as the values of a complete set of physical quantities that can measured on this system. This complete set of physical quantities is called a {\it context}, and a modality is attributed to a system within a context. {\color{black} Contexts are concretely defined by the settings of macroscopic measurement devices.}

\item[\bf P$_1$] {\bf Predictability and extravalence} --
Once a context is defined {\color{black} and the system is prepared in this context}, modalities can ideally be predicted with certainty and measured repeatedly on this system. {\color{black} When changing the context, modalities change but s}ome modalities in different contexts may be connected with certainty, this is called extracontextuality. This defines an equivalence class between modalities, called extravalence \cite{trsa,csm4c}.

\item[\bf P$_2$]  {\bf Contextual  quantisation}  --  For a given system and context, there exist at most $D$ distinguishable modalities, that are mutually exclusive: if one modality is realised in an experiment yielding a result in the macroscopic world, the other ones are not realised. 
The value of $D$, called the dimension, is a characteristic property of the quantum system, and is the same in all relevant contexts.

\item[\bf P$_3$] {\bf Changing context} -- 
Given P$_1$ and P$_2$, the different contexts relative to a given quantum system are related between themselves by continuous transformations (e.g. rotating a polarisation beamsplitter) which are associative, have a neutral element (no change), and an inverse. Therefore the set of context transformations has the structure of a continuous group, generally non commutative.
\end{itemize} 

For the sake of clarity, let us note that, within the usual QM formalism (which is not here yet),  a complete set of commuting observables (CSCO) corresponds to a context,  and a state vector (or projector onto that vector) corresponds to an  extravalence class of modalities, but {\bf not} to a particular modality, since  the specification of the context is missing in the state vector. 

The crucial postulate P$_2$  (Contextual  quantisation)  can be understood as the consequence of dealing with the smallest bits of reality, that for this reason have only a finite quantity of information to give whatever the way they are interrogated \cite{Rovelli}. This is also  related  to the existence of truly indiscernable microscopic objects:  if there were an infinite quantity of information that could be carried by individual quantum objects, they would end up all different at least on one of these pieces of information -- but both theoretical and empirical evidence tells us that QM does not work that way.
%\\

Then the leading idea of CSM is to start from these physical postulates, that involve the quantisation of the properties of microscopic systems within macroscopic contexts, and to show first that a probabilistic description is required to avoid contradiction \cite{csm1}. 
 The basic idea is that the results from different contexts cannot be gathered together, as this would lead to more than $D$ mutually exclusive modalities, contradicting P$_2$. Therefore the link between modalities in different contexts {\it must be probabilistic:} given a modality in a first context, only the probability to get another modality in a different context can be predicted \cite{random}. {\color{black} Additionally, it appears that this probabilistic aspect and its underlying indeterminism are key to maintain relativistic causality when spatially extended contexts are considered. As a matter of facts thanks to them, only randomness – i.e. non-information, entropy – can be transferred over space-like intervals. This is related to the discussion on predictive incompleteness presented in \cite{inference}, see e.g. Fig. 1 in this reference. }

It can also be shown that usual classical probabilities are not suitable to warrant that (i) there is a fixed number of mutually exclusive modalities in any context, and (ii) that the certainty of extravalent modalities can be transferred between contexts; in TBQM this is most directly shown by the Kochen-Specker theorem \cite{KSreview,KS}. A more suitable framework (worth trying !) is then to associate mutually orthogonal projectors to the events corresponding to mutually exclusive modalities \cite{csm4b,svozil}. 

{\color{black} 
\begin{itemize}
\item[\bf P$_4$]{\bf Projective probabilities} -- In a given context $\mathcal{C}$, each exclusive modality $\{m_i\}_{i=1}^D$ of a system is represented by a projector $\hat \Pi_i$ %on a direction 
in a Hilbert space of dimension $D$, all $\hat \Pi_i$'s being orthogonal. 
%\\
%\\
\end{itemize}

Then the above postulates can be used to justify the hypotheses of powerful mathematical theorems, that are respectively  Uhlhorn's theorem to connect different contexts with unitary transformations, and Gleason's theorem {\color{black}to derive} Born’s rule \cite{csm4c}. 
The  projectors are thus the basic mathematical tools, related to probabilities, and from them it is easy to construct standard observables:

\begin{itemize}
\item[\bf D$_2$]{\bf Observables as operators} -- 
From the orthogonal  rays given by the $\hat \Pi_i$'s, hermitian operators on a $D$-dimensional Hilbert space can be constructed by considering each  $\hat \Pi_i$ as an eigenspace associated to $m_i$, the corresponding eigenvalue. If $m_i$ corresponds to a single observable quantity, this yields an operator $ \hat M = \sum_{i=1}^D m_i \hat \Pi_i $. If $m_i$ is a tuple of several observable quantities, a tuple of operators can be constructed in a similar way. 
\end{itemize}

One thus recovers the standard definition of observables in a complete set of commuting observables (CSCO), in relation with the spectral theorem.} {\color{black} In this framework, traces over products of projectors and observables thus emerge naturally as the way to compute an experimental expectation value of the said observable. } 

Again, the intuitive idea behind the postulates is that making more measurements in QM (by changing the context) cannot  provide more details about the system, because this would increase the number of mutually exclusive modalities, contradicting P$_2$. One might conclude that changing context totally randomises all results and that nothing can be predicted, but this is not true either: some modalities may be related with certainty between different contexts, this is why extravalence is an essential feature of the construction {\color{black} -- actually, extravalent modalities tie the other, non extravalent modalities to a predictible probability distribution, through Gleason's theorem \cite{csm4c,csm4b}.} 
\\

\begin{figure}[h]
%\begin{minipage}{22pc}
\begin{center}
\includegraphics[width=0.85 \columnwidth]{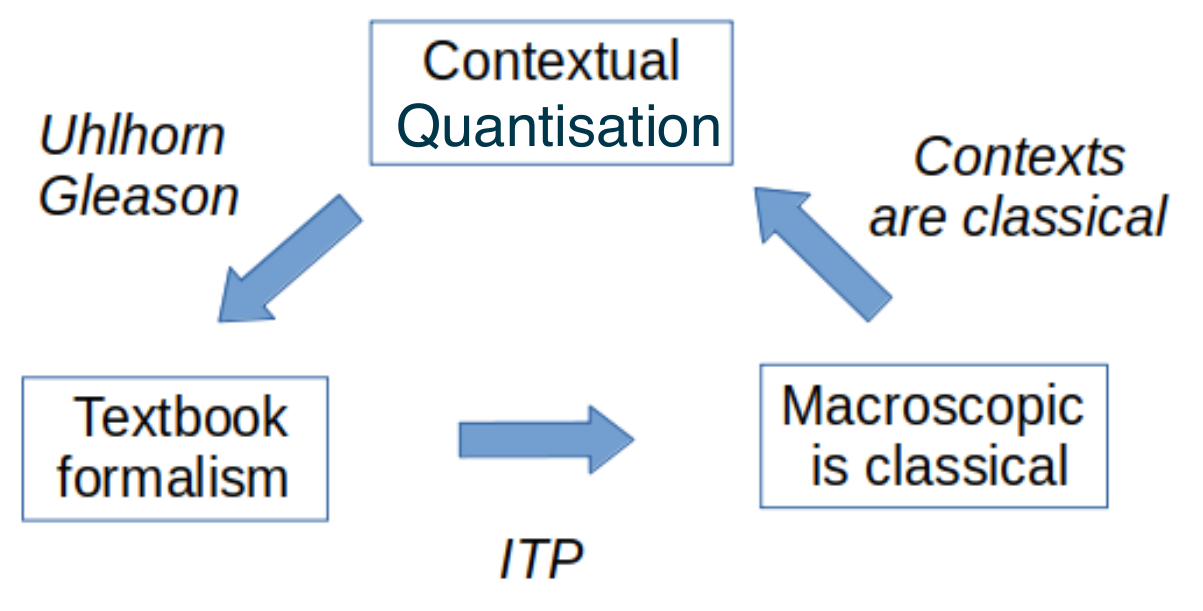}
\caption{\label{selfConsistency} 
Closing the loop of  Contexts, Systems and Modalities (CSM) by using Infinite Tensor Products (ITP).}
\end{center}
\end{figure}

The final step, moving out from usual TBQM, is to apply this formalism to a countable infinity of systems with infinite tensor products (ITP) of elementary Hilbert spaces, to model the macroscopic limit. 
{\color{black} Doing so, one moves from the separable Hilbert spaces and type-I operators considered in TBQM towards non-separable Hilbert spaces and Type-II and type-III operators  described by Murray and von Neuman in the 1930's and 1940's \cite{onRingsOfOperators}. 
In this limit}, unitary equivalence and unitarity overall are lost \cite{emch}, and the predicted behaviour looks very much like the classical one \cite{completing,FoP2023,DICE2024,Earman,GCS-IV}. Therefore the overall mathematical description fits with the initially postulated separation between microscopic systems and macroscopic contexts (Heisenberg cut), closing the loop of the reconstruction (Fig. \ref{selfConsistency}). 
\\

%%\vspace{-1cm}
\subsection{Discussion.}

We note that in the above approach there is no need to call for partial traces, or loss of information, since decoherence is built in initially by the postulates, and finally recovered from the (mathematically) infinite character of the context; this full loop is thus self-consistent. It is also quite possible to make type-I  calculations, for instance to calculate decoherence times in a given experiment; but it should be made explicit that they are approximations, able to get very close to the actual non-unitary jump, but unable to manage it. On the other hand, the overall framework sets a clear separation between the microscopic (system) level and the macroscopic (context) level, and makes sure that there is nothing like super-contexts, or some variety of Wigner’s friend, that would be able to turn an unbounded context back into a system \cite{extWigner}. Similarly, reasonings based upon a universally extended unitary evolution do not fit in our framework. 
%\vspace{-3mm}
\subsection{The crucial role of unitary transformations.}
%\vspace{-3mm}
Given the above statements, it is important to give more details on what unitary transformations are, and are not, 
according to the CSM approach. In standard QM, unitary transformations appear with a variety of different roles. A standard one is time evolution -- on which we will come back below. In relation with the previous sections, one may look at a related issue, that is the role of unitary transformations  in quantum measurements. This turns out to be quite important in the CSM framework, since a modality must be associated with a certain and repeatable result in a given context. However, this is not the most frequent situation in practical QM: actually, in most cases a unitary transformation is inserted in the quantum measurement itself. 

For instance, let us ask how to make a measurement that gives  a certain and repeatable modality, in the following situations: (i) a coherent state $|\alpha \rangle$ of an harmonic oscillator (or quantized electromagnetic field mode), (ii) a Bell state for two spins, and (iii) an arbitrary state of a quantum register.  Looking first at a coherent state $|\alpha \rangle$, it is clear that neither photon counting nor coherent detection will do the job: the first one gives a Poisson distribution of photo-counts, and the second one gives an amplitude value, with some probability distribution depending on how the measurement is implemented. But how to get the required certainty can be guessed easily \cite{CO2002}: let us deterministically translate $|\alpha \rangle$ by $(- \alpha)$, get the vacuum $|0 \rangle$, count zero photon with certainty, and translate back by $(+ \alpha)$ to the initial state. The modality criterion is thus respected, but it is clear that the irreversible part of the measurement (the photon counting) is inserted between two reversible unitary transformations, here translations. 

A similar situation appears to carry out a measurement in the Bell state basis, with the four entangled states $\{ |++ \rangle \pm |- - \rangle, |+- \rangle \pm |- + \rangle \}$. One has then to use a CNOT gate, then a Hadamard gate, measure the  spins along $z$ 
in the factorized basis $\{ |++ \rangle,  |- - \rangle, |+- \rangle ,  |- + \rangle \}$, and go back to the Bell basis by using the reverse unitary transform. This can obviously be generalized to an arbitrary quantum state of a register with many qubits: if such a state is prepared from the unitary transform $\hat U$, then apply $\hat U^\dagger$, check that the register is back to its initial state (all zeros for instance), and get back to the arbitrary quantum state by applying $\hat U$ again. 
%\\

Obviously these examples are highly idealized, since they assume perfect unitary transforms, and perfect quantum non demolition (QND) measurements when the irreversible check is performed. They are however perfectly legitimate from a quantum point of view, and match actually how a perfect quantum computer should be working. The current gate fidelities make that the successive application of $\hat U$ and $\hat U^\dagger$ cannot efficiently bring back to the initial state, unless a very small number of qubits is used -- doing it with a very large register is extremely challenging, though not forbidden in principle.

In the logic of CSM these examples make clear that unitary transformations describe the deterministic evolution or manipulation of isolated quantum systems within classical contexts, outside the measurement periods \cite{completing,FoP2023}. However, they don't apply to the universe as a whole. The contexts themselves are classically described, and do not correspond to  mathematical entities that can be the object of unitary transformations; quite the opposite, in an algebraic framework they correspond to separate sectors, that are not connected by any operator constructed from the systems level \cite{DICE2024}. This illustrates again the overall consistency of CSM, from the previous physical postulates to the mathematical formalism, and back. 
%%%%%%%%%%%%%%%%%%%%%%%%%%%%%%%%%%%%%%%%%%
\section{Higher level implications}
The scheme summarized above builds the TBQM formalism from a few postulates and then closes the loop by showing how the ITP limit recovers the key assumptions. This construction  calls examination from a higher level perspective, at least on three aspects (i) the acceptability of the infinite system limit (ii) the key role of unitarity and where it cannot apply and (iii) a new light on the debate between those who consider that classical physics should emerge from quantum (reductionism), and those who think that there is a fundamental distinction (dualism) -- actually both positions might be {\it equivalent}. We discuss these three aspects in this section.
\vspace{-3mm}
\subsection{Is infinity acceptable at all?}
%\vspace{-3mm}
Using infinity in a physical theory legitimately rises relevant questions. These questions relate intuitively to the one whether our universe is infinite or not -- which answer we do not know. Moreover, in the specific case we are considering, something even more puzzling occurs. At the infinite subsystem number $N\rightarrow \infty$ limit, the ITP of $N$ elementary Hilbert space $\mathcal{H}=\otimes_{\alpha=1}^N  \mathcal{H}_\alpha$ appears to become suddenly non-separable, i.e. qualitatively different, which could lead to thinking that anything valid at $N < \infty$ does not hold anymore in the limit. Yet, there are two points one can make to justify taking the limit seriously. One is mathematical, the other one is more epistemological.

{\it On the mathematical side}, looking at the details of von Neumann's breakdown theorem \cite{FoP2023,DICE2024}, things are much more subtle than just a function that would be discontinuous at the  limit. As a matter of fact, two key properties of the non-separable ITP Hilbert space are
  
\noindent {\bf (1)} the breakdown into {\color{black}non-unitarily equivalent} orthogonal sectors that correspond to an infinite number of changes in the elementary subsystem states, and
 
\noindent {\bf (2)} the fact that sectors are not connected by the ring $\mathcal{B}^\#$ of operators, built as the extension to the full ITP of operators that act on elementary subsystem Hilbert spaces, their products, sums, and topological completions.
%\vspace{1mm}

%\noindent 
\noindent Quite importantly, it can be shown that these two properties build up gradually. More precisely \cite{JvN39} shows that 

\noindent {\bf (1)} if  $\vert \Psi\rangle := \otimes_{\alpha=1}^N\vert \psi_\alpha\rangle $ and $\vert \Phi\rangle := \otimes_{\alpha=1}^N\vert \phi_\alpha\rangle $ in $\mathcal{H}$ are not in the same sector when $N \rightarrow \infty$, for any $\varepsilon > 0$ one can find a finite set $J\subset [1,...,N]$ of $M$ indices $\alpha$'s, all distinct, so as to build $\vert \Psi_M\rangle := \otimes_{\alpha\in J}\vert \psi_\alpha\rangle $ and $\vert \Phi_M\rangle := \otimes_{\alpha\in J}\vert \phi_\alpha\rangle$ such that $\vert \langle \Psi_M \vert \Phi_M \rangle \vert < \varepsilon$

\noindent  {\bf (2)} assume $\hat A$ is a bounded operator in $\mathcal{B}^\#$. 
If  $\vert \Psi \rangle$ and $\vert \Phi \rangle$ are not in the same sector when $N \rightarrow \infty$, for any $\varepsilon > 0$ 
one can find a finite set $J\subset [1,...,N]$ of $M$ indices $\alpha$'s, all distinct, so as to build 
 $\vert \Psi_M\rangle$ and $\vert \Phi_M\rangle$ as above, and the restriction $\hat A_M$ of $\hat A$ to the $\otimes_{\alpha \in J}\mathcal{H}_\alpha$, such that
$\vert \langle \Psi_M \vert \hat A_M\vert\Phi_M \rangle \vert < \varepsilon$.  
%\vspace{1mm}

%\noindent  
 This gradual onset of the properties means that the non-separable, brokendown limit is reached in a controlled manner, at least in the weak topology that is the one relevant for von Neumann ($W^*$-)algebras\footnote{Note that we make no claim here about $C^*$-algebras that rely on a different topology to control the limits, because as described above, our quantum formalism needs projectors and traces, the latter not being necessarily available in $C^*$-algebras}. This controlled approach to the limit is very much reminiscent of the controlled approach to the Central Limit Theorem at the Thermodynamical limit, on which much of equilibrium statistical mechanics relies.
Another example is the pervasive function derivative,  that consider infinitesimal elements even though we might think that there is also an ultraviolet cutoff at the Planck scale that make them no more valid than the thermodynamic limit. On top of this, this gradual onset can be understood in a `for all practical purposes' way, in the sense that for a large enough system, the inter-sector coherences in the density operator are so weak that it would take experimental repetitions over more than the age of the Universe to observe a quantum effect in such a system.   

{\it From the epistemological point of view}, this limit can be validated too. As a matter of fact, however generic a conceptual representation of reality may be, it remains a {\it model} of reality \cite{SciRepStanford}. 
Generally,  physicists assume that: 

\noindent $\bullet$ there is a mapping between concepts in the representation (that can be expressed in mathematical language), and the target elements of reality %and 

\noindent $\bullet$ this mapping allows conducting surrogative reasoning \cite{swoyer} 
on the concepts to yield (falsifiable) claims on the elements of reality they are meant to describe. 

%\noindent 
Though this might be blurred in the daily exercise of physics, representations and reality are elements of two different worlds, thus the conceptual elements of a model are not elements of reality. Moreover, models are by definition approximate, valid until proven wrong by an experiment. So models do not need to have all the properties of reality to be relevant, especially in the most remote corners of their application domain\footnote{However, the history of science has shown that unexpected properties of representations could actually have surprising counterparts in reality.}. 

Overall we consider that these two arguments in two distinct domains validate the consequences which can be derived from taking the $N\rightarrow \infty$ limit in QM.

\vspace{-3mm}
\subsection{Unitarity relevance and multiverse interpretation}
%\vspace{-3mm}
We come back here to a standard role of unitary transformations in QM, that is time evolution. Assuming a Hamiltonian $\hat H(t)$ that describes the energy of an otherwise isolated system, it is well known that the system's evolution can be described by a unitary operator $\hat U(t)$, solution of the equation $i \hbar \; d\hat U(t)/dt = \hat H(t) \hat U(t)$. 
More generally, when a transformation (rotation, translation...) is applied at the classical level, one can define a corresponding unitary transformation to be applied on the states or observables of the system. In this point of view, time evolution is just a translation in time, and the Hamiltonian is the infinitesimal generator of such translations.  
This important subject can be developed in great details, and it shows the importance of Uhlhorn's theorem to build  representations of symmetry groups \cite{bookFL}.

This role of unitarity in time evolution, contrasted with the abrupt evolution during a measurement, is a disturbing situation taking into account the tremendous success of QM. 
All the more when consideration{\color{black}s} on measurement{\color{black}s} are extended to the whole Universe, where it would lead to the `many worlds' conclusion that there is an infinity of parallel different universes, where each possible outcome of any measurement is realised.  
In our approach spelled out above, these extrapolations are unwarranted, and result from a misunderstanding and misuse of the quantum formalism.  

Nevertheless, in defense of the idea of multiple parallel universes, it may be told that Science already made many counter-intuitive predictions, establishing e.g.  that the earth is round and moving quite fast, despite the  `obvious empirical evidence’ that it is flat and motionless. 
But actually this view raises two issues of different natures. 

\noindent $\bullet$ For it to be a scientific statement, it would require to yield a falsifiable experimental prediction -- analog to Bell's inequalities for local hidden variables. Such a prediction is not available yet;
actually, this idea only shows up as a consequence of extrapolating the {\color{black}type-I} quantum formalism, by applying {\color{black}carelessly} it to macroscopic systems and then to the whole universe. This is at difference with the round and moving aspects of the Earth, which led immediately to many practical predictions, e.g. sailing around it, that have been largely vindicated. 
 
\noindent $\bullet$ The above considerations on ITP show (if the model holds) that there is no reason to expect any unitarity whatsoever at a macrocopic scale, and thus that the very motivation for parallel universes collapses.

But all this is not a real surprise, as the previous section explained that TBQM can be derived from a set of postulates that include the unicity of the Universe.
\vspace{-3mm}
\subsection{Reductionism vs dualism}
%\vspace{-3mm}
Despite all classical objects being built with of quantum objects, the radical difference between the 
classical and quantum behaviours 
has raised very early in the history of quantum physics the question of their compatibility. Two antagonistic positions have emerged. Bohr and followers have claimed that there are two fundamentally different levels of reality, separated by a `Heisenberg cut' and that physicists have to live with this dual description of reality. Let us call `dualist' this position.  The other position aims at looking for a mechanism, by which classical behaviours would emerge from quantum ones. One could see Ehrenfest's theorem or intense coherent states \cite{Paola} as first hints in this direction, that has then been much developed in the frame of decoherence theory \cite{zeh,zurek}. Classical would reduce to a part of quantum, in this `reductionist' perspective.    
\vspace{3 mm}

The closing of the CSM loop with infinite tensor products sheds a new light on this antagonism. 
In the language of CSM, the dualist approach translates into considering that there is always a (classical) context around a quantum system, with a Heisenberg cut separating them. This assumption is the prerequisite of the contextual quantisation postulate [II.B]. These postulates allow deriving usual TBQM, so duality and contextuality {\color{black} lead to} QM. But now, forgetting that this usual QM formalism can result from dualism, and just taking it for granted, one sees that key properties of contextuality (the Kochen-Specker theorem) and the difference between quantum and classical behaviours (through von Neumann's breakdown theorem on ITP) result from usual QM. So QM implies duality and contextuality. In other words, dualism implies reductionism and reductionism implies dualism. In logical terms, this means that both positions can be viewed as {\it equivalent}, and not antagonistic.   
%\\
%%\vspace{-3mm}

%%%%%%%%%%%%%%%%%%%%%%%%%%%%%%%%%%%%%%%%%%

\section{Conclusion: QM for engineers, and beyond ?  }
%\vspace{-2mm}
There are now many engineers working on quantum technologies, and for engineering it is clear that a reliable physical ontology is extremely useful, to tell which objects and which properties they are working with. In such a framework, speaking about inaccessible multiple worlds or dead-and-alive cats is not very enlightening; invoking {\color{black}only} abstract equations is not much better. 

So coming back within our unique world, for the worst and for the best, may be quite useful in practice. 
Also, rather than telling that a quantum superposition is ‘being in two states at the same time’, it is better to tell that one can get a result with certainty for some measurement in some context, and a random result for some other measurements in another context {\color{black} determined at measurement time}. Maybe the strangest feature of quantum randomness is that it can be turned into a certainty -- for well chosen measurements. But when seen from the engineering side, this is a quite manageable idea, likely to orient thinking in  {\color{black} a practically usable} direction. For simple accounts see also \cite{myst,debate}. 
%\\

From a more foundational perspective, it is clear that the views presented here have a strong Bohrian flavour, though they are quite distinct from Bohr’s ideas; for instance we never speak about complementarity, that is too vague in our opinion. Also, it can be said that our approach is close to the so-called {\color{black}Copenhagen} point of view -- certainly closer to it than to any other `interpretation' of QM. However, {\color{black}the Copenhagen} framework is also loosely defined, and it does not include topics like non-type I operator algebra that are essential for our construction. {\color{black}As a matter of facts, considering} the use of non-type-I algebras allows proposing a global model where the classical and the quantum realms have a clearly articulated relationship, clarifying the way each one 
relates to the other.

So overall we call for an extension of the QM formalism towards operator algebras, despite the known mathematical difficulties of this topic.  But maybe this field went too quickly to mathematics, so it should be reconsidered by physicists, and brought back into their realm. 
\vspace{-2mm}

\acknowledgments{PG thanks Franck  Lalo\"e,  Roger Balian and Olivier Ezratty for many interesting discussions. }


\begin{thebibliography}{999}


\bibitem{FL}  Lalo\"e, F. {\it Do We Really Understand Quantum Mechanics?}, Cambridge University Press (2012).

\bibitem{JvN39} von Neumann, J. On infinite direct products,  {\it Compositio Mathematica 6}, 1-77 (1939). \\
http://www.numdam.org/item?id=CM$\_1939\_\_6\_\_1\_0$

\bibitem{onRingsOfOperators} Murray, F. J. and von Neumann, J. On Rings of Operators IV, {\it Ann. Mathematics  44}, 716 (1943) and references therein.

\bibitem{emch}  Emch, Gerard G. {\it Algebraic Methods in Statistical Mechanics and Quantum Field Theory}, 
Dover, New York, 2000 (reprint of the Wiley-Interscience 1972 edition).

\bibitem{CCT} Cohen-Tannoudji C., Diu B., Lalo\"e F.  {\it  Quantum mechanics} (3 volumes), John Wiley \& Sons (1977-2019)

\bibitem{zeh} Zeh, H.D.  On the Interpretation of Measurement in Quantum Theory, 
{\it Found. Phys., 1}, 69–76, (1970).


\bibitem{zurek}  Zurek, W.H. (2022). Emergence of the Classical World from Within Our Quantum Universe. In: Kiefer, C. (eds) {\it From Quantum to Classical. Fundamental Theories of Physics, vol 204.} Springer, Cham.  [arxiv:2107.03378].


\bibitem{csm4c}  Auff\`eves, A. and Grangier, P.  Revisiting Born's rule through Uhlhorn's and Gleason's theorems, 
{\it Entropy 24 (2)}, 199 (2022) https://doi.org/10.3390/e24020199
%https://www.mdpi.com/1099-4300/24/2/199
%https://arxiv.org/abs/2111.10758

\bibitem{CO2002} Grangier, P. Contextual objectivity: a realistic interpretation of quantum mechanics, {\it European Journal of Physics 23:3}, 331 (2002) [arXiv:quant-ph/0012122]. 

\bibitem{csm1} Auff\`eves, A. and Grangier, P. Contexts, Systems and Modalities: a new ontology for quantum mechanics, {\it Found. Phys. 46}, 121 (2016) [arXiv:1409.2120].
%\\

\bibitem{trsa} Auff\`eves, A. and Grangier, P. Extracontextuality and extravalence in quantum mechanics, 
{\it Phil. Trans. R. Soc. A 376}, 20170311 (2018) [arXiv:1801.01398].


\bibitem{random} Grangier, P. and Auff\`eves, A. What is quantum in quantum randomness?, 
{\it Phil. Trans. R. Soc. A 376}, 20170322 (2018) [arXiv:1804.04807].

\bibitem{myst}  Grangier, P. Revisiting quantum mysteries, in {\it The Quantum-Like Revolution: A Festschrift for Andrei Khrennikov}, A. Plotnitsky and E. Haven eds, Springer Cham (2023) [arxiv:2105.14448].


\bibitem{csm4b}  Auff\`eves, A. and Grangier, P. Deriving Born's rule from an Inference to the Best Explanation, 
{\it Found. Phys. 50, 1781-1793} (2020) [arXiv:1910.13738].  
%\\  https://doi.org/10.1007/s10701-020-00326-8, 

\bibitem{inference}  Grangier, P. Contextual inferences, nonlocality, and the incompleteness of quantum mechanics, 
 {\it Entropy 23 (12)}, 1660 (2021); https://doi.org/10.3390/e23121660 
%http://arxiv.org/abs/2012.09736 %[quant-ph]  (2020). 

\bibitem{debate}  Farouki, N. and Grangier, P.  The Einstein-Bohr debate: finding a common ground of understanding ?, 
{\it Found. Sci. 26}, 97-101 (2021)
%https://doi.org/10.1007/s10699-020-09716-7
[arXiv:1907.11267].

\bibitem{KS} Grangier, P. Revisiting Quantum Contextuality, https://arxiv.org/abs/2201.00371 (2022)

\bibitem{completing}  Grangier, P. Completing the quantum formalism in a contextually objective framework,  
{\it Found. Phys.  51}, 76 (2021) [arXiv:2003.03121].

\bibitem{FoP2023}   
Van Den Bossche, M.  and Grangier, P. Contextual unification of classical and quantum physics, 
%
{\it Found. Phys.}  53:45 (2023) [arXiv:2209.01463]. 

\bibitem{DICE2024}
Van Den Bossche, M.  and Grangier, P. Revisiting Quantum Contextuality in an Algebraic Framework,
%https://arxiv.org/abs/2304.07757
{\it J. Phys.: Conf. Ser. 2533} 012008 (2023), %Journal of Physics: Conference Series
 Proceedings of the DICE 2022 Conference, Castiglioncello, Italy. [arxiv:2304.07757]
% Journal of Physics: Conference Series, Volume 2533, Tenth International Workshop DICE2022 - 
% Spacetime - Matter - Quantum Mechanics 19/09/2022 - 23/09/2022 Castiglioncello, Italy 
 
\bibitem{extWigner} Federico, M.  and Grangier, P.  A contextually objective approach to the extended Wigner's friend thought experiment, https://arxiv.org/abs/2301.03016 (2023)
\\
\\

\bibitem{svozil} Svozil, K. Extending Kolmogorov's axioms for a generalized probability theory on collections of contexts 
%https://arxiv.org/abs/1903.10424
[arxiv:1903.10424].

\bibitem{Rovelli} Rovelli, C., Relational quantum mechanics, {\it Int. J. Theor. Phys. 35} 1637  (1996)
[arXiv:quant-ph/9609002].

\bibitem{KSreview} Budroni, C., Cabello, A., Gühne, O., Kleinmann, M., Larsson, J.-A.,   Kochen-Specker contextuality, {\it Rev. Mod. Phys.  94:4} 045007 (2022) [arXiv:2102.13036].

\bibitem{Earman}  Earman, J. Quantum Physics in Non-Separable Hilbert Spaces, http://philsci-archive.pitt.edu/18363/
This interesting analysis agrees with us on the mathematical side, though not on the physical side. 

\bibitem{GCS-IV} Thiemann, T., Winkler, O. Gauge field theory coherent states: IV. Infinite tensor product and thermodynamical limit,  {\it Class. Quantum Grav. 18}, 4997 (2001) [arXiv:hep-th/0005235]. %DOI 10.1088/0264-9381/18/23/302
%Though its purpose is quite different from ours, 
This article (sec. 4) gives an interesting presentation and discussion of \cite{JvN39}.

\bibitem{SciRepStanford} See e.g. https://plato.stanford.edu/entries/scientific-representation/

\bibitem{swoyer} Swoyer, C. Structural representation and surrogative reasoning, 
{\it Synthese 87}, 449-508 (1991)

\bibitem{bookFL}  Lalo\"e, F.  {\it Sym\' etries continues.} 
Collection Savoirs actuels - Math\' ematiques, EDP Sciences (2021).
%ISBN 978-2-7598-2631-5

\bibitem{Paola} Coppo, A., Pranzini, N., Verrucchi, P. 
Threshold size for the emergence of classical-like behavior {\it Phys. Rev. A 106} 042208 (2022) [arxiv:2203.13587].

\end{thebibliography}
\end{document}